\documentclass[twocolumn,showpacs,showkeys,preprintnumbers,prd,nofootinbib,superscriptaddress]{revtex4-1}

\usepackage{subfigure}
\usepackage{amsfonts,amsmath,amssymb}
\usepackage{graphicx}
\usepackage{inputenc} 
\usepackage{hyperref}
\usepackage{color}

\begin{document}

\title{Constraining primordial black holes as a fraction of dark matter through accretion disk luminosity}

\author{Rocco D'Agostino}
\email{rocco.dagostino@unina.it}
\affiliation{Scuola Superiore Meridionale, Largo S. Marcellino 10, 80138 Napoli, Italy.}
\affiliation{Istituto Nazionale di Fisica Nucleare (INFN), Sez. di Napoli, Via Cinthia 9, 80126 Napoli, Italy.}

\author{Roberto Giamb\`o}
\email{roberto.giambo@unicam.it}
\affiliation{School of Science and Technology, University of Camerino, Via Madonna delle Carceri 9, 62032 Camerino, Italy.}
\affiliation{Istituto Nazionale di Fisica Nucleare (INFN), Sezione di Perugia, Via Alessandro Pascoli 23c, 06123 Perugia, Italy.}

\author{Orlando Luongo}
\email{orlando.luongo@unicam.it}
\affiliation{Dipartimento di Matematica, Universit\`a di Pisa, Largo B. Pontecorvo, 56127 Pisa, Italy.}
\affiliation{Universit\`a di Camerino, Divisione di Fisica, Via Madonna delle carceri, 62032 Camerino, Italy.}
\affiliation{NNLOT, Al-Farabi Kazakh National University, Al-Farabi av. 71, 050040 Almaty, Kazakhstan.}

\begin{abstract}
In this paper, we consider the hypothesis that fractions of dark matter could be constituted by primordial black holes (PBHs). To test this possibility, we work out the observational properties of a static black hole embedded in the dark matter envelope made of a PBH source. The corresponding modifications of geometry due to such a physical system are investigated, with a particular focus on the accretion disk luminosity in spiral galaxies. 
The impact of the PBH presence is analyzed through modification of the disk luminosity and kinematic quantities. Thus, we discuss possible constraints on the PBH abundance in view of the most recent theoretical bounds. The results of our study indicate that suitable PBH masses are $M_\text{PBH}\in[10^6,10^{12}]M_\odot$ for PBH fractions   $f_\text{PBH}\in[10^{-3},1]$. In particular, a comparison with the predictions of the exponential sphere density profile for dark matter suggests that the best-matching configuration is achieved for $f_\text{PBH}=1$ and $M_\text{PBH}=10^6 M_\odot$.
Consequences with respect to the current knowledge on primordial black hole physics are discussed. 
\end{abstract}


\maketitle

\section{Introduction}
\label{sec:1}

Last year's observations certify that we are currently encompassing the epoch of black hole astronomy \cite{Bambi:2019xzp}. 
The first evidence for  supermassive black hole\footnote{Theoretical scenarios for BH formation and distribution fail to be 
predictive \cite{Kormendy:2013dxa}, i.e.,  their formation is far less understood with respect to their light, stellar 
mass, and counterparts.} (BH) shadow in the galaxy M87 \cite{EventHorizonTelescope:2019dse,EventHorizonTelescope:2019ggy} has shed light on 
galaxy formation and on the presence of extreme compact objects located at the centers of spirals \cite{Cattaneo:2009ub}.
After the first detection of gravitational waves \cite{LIGOScientific:2016aoc}, BH and compact object mergers opened new windows towards the so-called \emph{multi-messenger
astronomy} \cite{Meszaros:2019xej}. The most accredited strategy to infer the supermassive BH masses handles measures of accretion disk spectra\footnote{A significant exception is offered by Sgr-A* in the Milky Way and the supermassive black hole candidate in the galaxy M87.} \cite{Abramowicz:1988sp}. An accretion disk, determined by a central highly-massive object, represents a disk-like flow configuration constituted by fluids under a particular physical state, such as gas, plasma or simply  dust. The role played by  pressure cannot be avoided, if one considers cases that do not include dust only. In fact, even particles around a massive astronomical object may represent the material orbiting in the gravitational field providing the accretion disk itself. This configuration typically requires that the material configuration loses energy and angular momentum as it slowly spirals inward.

The accretion disk properties turn out to be essential in order to predict theoretical models that, regardless of the morphology of galaxies, are able to place constraints on general relativity and/or extended theories of gravity \cite{Capozziello:2019cav,DAgostino:2020dhv,DAgostino:2022tdk}. 
In this respect, the underlying geometry plays a central role  \cite{Abramowicz:2011xu}. Considering accretors at the center of a given galaxy requires the presence of dark matter (DM) to fuel the material orbiting within the accretion disk. It is then reasonable to believe that the central distribution may affect the overall geometry, once the DM distribution is somehow known \cite{Salucci:2018hqu}. In analogy, relativistic effects might be included in describing the DM distribution, modifying consequently the light properties  emitted by accretion disks. 

An intriguing possibility is offered by considering DM under the form of PBHs \cite{Green:2020jor}. 
Strong observational evidence suggests that the matter content of the Universe is mostly made of non-baryonic particles under the form of cold DM (see, for a review, \cite{Bertone:2016nfn}).
After the earlier works studying the formation of PBHs from the gravitational collapse in the primordial Universe \cite{Zeldovich:1967lct,Hawking:1971ei}, it was soon realized that a potential candidate for DM may be represented by PBHs \cite{Chapline:1975ojl}. Since their formation dates back to before  matter-radiation equality, PBHs are characterized by a non-baryonic nature and could, in fact, behave like DM particles. PBHs endowed with an initial mass $M_\text{PBH}\gtrsim 10^{14}$ g are thought to escape Hawking's evaporation and live longer than the Universe has existed \cite{Page:1976df,MacGibbon:2007yq}. 

A surge of interest in PBHs as DM was brought on by the microlensing results of the MACHO Collaboration \cite{MACHO:1996qam}, which observed a much greater number of events with respect to those expected from stellar populations. Such excess is consistent by considering that half of the halo of our galaxy is made of compact objects with mass $0.5M_\odot$, while considerations based on the baryon budget would exclude astrophysical compact objects \cite{Fields:1999ar}.

A more recent wave of interest in considering DM made of PBHs was given by the detection of gravitational waves by the LIGO-Virgo Collaboration \cite{LIGOScientific:2016aoc}. In particular, it was argued that  the merging solar mass BHs responsible for the signal could have a primordial origin rather than astrophysical \cite{Bird:2016dcv,Clesse:2016vqa,Sasaki:2016jop}.

Moreover, this scenario has been reinforced due to the missing  evidence for the most popular DM candidates, such as WIMPs, sterile neutrinos and axions, although the huge experimental efforts of the last decades 
\cite{Bertone:2018krk,Bertone:2016nfn}. If we admit the existence of PBHs as plausible candidates for DM, it is not possible to exclude their presence around an accretor to contribute to the overall material distribution of the accretion disks of a given spiral galaxy. 

Motivated by these considerations, in this paper we assume that DM around a generic galaxy is made of a fraction of PBHs.  In particular, we study how the spectrum of the accretion disk, surrounding a highly-massive central compact object, could be modified according to the above hypothesis. Invaluable information about the accretion disk luminosity can be then obtained by varying the PBH abundance. To do so, we describe the geometry of a generic spiral galaxy in spherical symmetry. Hence, we investigate the radial contributions of density and pressure adopting a spherical Tolman-Oppeheimer-Volkov (TOV) spacetime, by means of suitable boundary conditions on the galaxy configuration. We thus work out the standard luminosity provided by the Novikov-Thorne approach. The flux and differential fluxes are computed, once the energy, angular velocity and momentum are inferred from the metric itself. Our analysis relies on the assumption that each DM particle is composed of PBH with the same mass, and that mass loss is negligible as well as angular momentum is assumed not to decrease. We evaluate the theoretical curves that will depend upon the PBH fraction. We thus put constraints over the expected PBH mass and abundance confronting our findings with the outcomes provided by the DM exponential sphere density model. 

The paper is structured as follows. In Section \ref{sezione2}, we characterize the PBH distribution in a DM halo. We highlight the main features of such a picture and underline their possible limitations. In Section \ref{sezione3}, we model DM in galaxies, pointing out the most suitable boundary conditions and indicative priors over the quantities under exam. In Section \ref{sezione4}, the accretion disk luminosity is computed by means of the Novikov-Thorne approach, which allows us to calculate both kinematic and spectroscopic quantities. Theoretical discussions have been reported in Section \ref{sezione5}, emphasizing the most suitable constraints on the PBH fractions that turn out to be compatible with the present expectations for DM. Finally, in Section \ref{sezione6} we report conclusions and perspectives.

\section{Primordial black holes as dark matter }\label{sezione2}

One of the possibilities explored for nearly fifty years to explain the production of DM in the early Universe is that the 
DM abundance observed today might be the relic from the evaporation of some populations of energetic enough PBHs. 
Remarkably, 
this scenario does not imply the existence of extra interactions, but it arises from particle production through Hawking 
radiation \cite{Hawking:1974rv}. 
Indeed, the collapse of BHs generates a thermal flux of particles that would constitute the source of gravitationally 
interacting DM \cite{Cheek:2021odj}. 
According to the standard cosmological scenario, the origin of the cosmic structures is due to matter perturbations seeded 
by an early period of exponential expansion of spacetime known as inflation\footnote{For a recent perspective, see 
\cite{DAgostino:2021vvv}.} \cite{Starobinsky:1980te,Guth:1980zm,Linde:1981mu}. Although the absence of a consensus scenario, the existence 
of PBHs is predicted by several inflationary models. 

The interest in the hypothesis that a significant fraction of DM might be made of PBHs has been recently renewed after the 
observations of gravitational waves from a merger of massive BHs detected by the LIGO/Virgo Collaboration 
\cite{LIGOScientific:2018mvr,LIGOScientific:2020stg,LIGOScientific:2020zkf}. 
The mass of PBHs has been constrained by a large variety of experiments over the years\footnote{For a comprehensive review, 
see e.g. \cite{Carr:2020xqk}.}. The minimum value is derived by requiring that they have not already evaporated. 
Specifically, from the Hawking radiation one finds that  $M_\text{PBH} \geq 5 \times 10^{14}$ g \cite{Carr:2020gox}.

To characterize PBH distribution in spiral galaxies, one can work out the simplest possible configuration, regardless of 
the redshift dependence and the two-point correlation function. This can be done by adopting the following assumptions: 
\emph{a)} all PBHs behave as copies of particles; \emph{b)} they do not interact with each other; \emph{c)} the volume in 
which they lie is exactly the spherical volume associated with the spacetime geometry; \emph{d)} the density of PBHs 
coincides 
with the DM density. 
Based on these simplistic assumptions, one could write the total DM mass as a function of the PBH mass density $\Phi(m)$ 
and the corresponding density as $M_\text{DM}=\mathcal N\int dm\,\Phi(m)$ and $\rho_\text{DM}=\frac{M_\text{DM}}{V}\propto 
r^{-3}$,
where the last proportionality represents the steepest condition for the radial dependence of the PBH density in terms of 
DM, and $\mathcal N$ is the number of PBHs.

Clearly, one may go beyond this prototype scheme and consider a more realistic scenario, which takes into account the 
redshift dependence and the two-point correlation function associated with PBHs to describe the radial distribution of the DM 
halo.

To this purpose, we follow the approach of \cite{DeLuca:2020jug} and consider the evolution of the PBH spatial distribution as a 
function of the redshift $z$ and of the comoving separation $x=|\vec{x}|$. In particular, the two-point correlation 
function 
of PBH is characterized by the overdensity of an individual PBH centered at the comoving position $\vec{x}_i$, given by
\begin{equation}
\dfrac{\delta\rho_\text{PBH}(\vec{x},z)}{f_\text{PBH}\overline{\rho}_\text{DM}}=\dfrac{1}{\overline{n}_\text{PBH}}\sum_i\delta_D(\vec{x}-\vec{x}_i(z))-1\,,
\end{equation}
where $\overline{\rho}_\text{DM}$ is the background DM energy density, and $\delta_D(\vec{x})$ is the 3-dimensional Dirac 
distribution. Here, $f_\text{PBH}\equiv\frac{\Omega_\text{PBH,0}}{\Omega_\text{DM,0}}$ is the fraction of DM under the form 
of PBHs, with $\Omega_\text{PBH,0}$ and $\Omega_\text{DM,0}$ being the density parameters of PBHs and DM at redshift $z=0$, 
respectively.

Also, $\overline{n}_\text{PBH}$ represents the PBH average number density per comoving volume:
\begin{equation}
\overline{n}_\text{PBH}\simeq 3.2\, 
f_\text{PBH}\left(\dfrac{20\,M_\odot/h_0}{M_\text{PBH}}\right)\left(\frac{h_0}{\text{kpc}}\right)^3\,,
\label{eq:n_medio}
\end{equation}
where $M_\text{PBH}$ and $M_\odot$ are the PBH and the solar masses, respectively, while $h_0\equiv 
H_0/(100\,\text{km}\,\text{s}^{-1}\text{Mpc}^{-1})$ is the reduced Hubble constant.
Therefore, one can write the PBH two-point correlation function as
\begin{equation}
\left\langle \dfrac{\delta\rho_\text{PBH}(\vec{x},z)}{\overline{\rho}_\text{DM}}\ 
\dfrac{\delta\rho_\text{PBH}(0,z)}{\overline{\rho}_\text{DM}}\right\rangle=\frac{f_\text{PBH}^2}{\overline{n}_\text{PBH}}\,\delta_D(\vec{x})+\xi(x,z)\,,
\end{equation}
where $\xi(x,z)$ is the reduced PBH correlation function. The above expression can be then used to define the PBH power 
spectrum relative to the total DM energy density:
\begin{equation}
\Delta^2(k,z)=\frac{k^3}{2\pi^2}\int d^3x\, e^{i\vec{k}\cdot \vec{x}}    \left\langle 
\dfrac{\delta\rho_\text{PBH}(\vec{x},z)}{\overline{\rho}_\text{DM}}\ 
\dfrac{\delta\rho_\text{PBH}(0,z)}{\overline{\rho}_\text{DM}}\right\rangle . 
\end{equation}

To proceed, we consider the Press-Schechter formalism \cite{Press:1973iz} applied to an initial Poisson power spectrum, 
from which it is possible to obtain the number density of PBH halos with a mass within the interval ($M,M+dM$):
\begin{equation}\label{fa}
\frac{dn(M,z)}{dM}=\frac{\overline{\rho}_\text{PBH}}{\sqrt{\pi}}\left(\frac{M}{M_\ast(z)}\right)^{1/2}\frac{e^{-M/M_\ast(z)}}{M^2}\,,
\end{equation}
where $\overline{\rho}_\text{PBH}=\overline{n}_\text{PBH}M_\text{PBH}$ is the average PBH energy density, and $M_\ast(z)$ 
is the typical halo mass that collapses at $z$ \cite{Hutsi:2019hlw}, which is given by
\begin{equation}
M_\ast(z)=N_\ast(z)\,M_\text{PBH}\simeq f_\text{PBH}^2\left(\frac{2600}{1+z}\right)^2M_\text{PBH}\,.
\end{equation}

In the framework of the halo model \cite{Sheth:1996ab,Cooray:2002dia}, the correlation function reads
\begin{equation}
\xi(r,z)=\frac{1}{\overline{\rho}_\text{DM}^2}\int dM\, \frac{dn(M,z)}{dM} M^2 \mu_M(r,z)\,,
\end{equation}
where 
\begin{align}
\mu_M(r,z)&=\int d^3s\, \rho_\text{PBH}(s,M,z)\,\rho_\text{PBH}(|\vec{s}+\vec{r}|,M,z) \nonumber \\
&\simeq \frac{1.22}{4\pi R_\text{vir}^3}\left(\frac{r}{R_\text{vir}}\right)^{-9/5}.
\end{align}
One thus obtains the PBH halo density profile as
\begin{equation}
\rho_\text{PBH}(r)= \frac{3M_\ast}{20\pi}R_\ast^{-3/5}r^{-12/5},
\label{eq:PBH_density}
\end{equation}
where $R_\ast$ is the virialized radius of a halo of mass $M_\ast$:
\begin{equation}
R_\ast=\left(\frac{3M_\ast}{4\pi\cdot200\,\overline{\rho}_\text{PBH}}\right)^{1/3},
\end{equation}
in which an average density of 200 times the background density is assumed within each virialized halo.

\section{Modelling dark matter in the galaxy}\label{sezione3}

The nature and the high mass of the accreting central object require the use of general relativity. The accretion disk luminosity will be therefore determined once the metric is involved as a solution to Einstein's field equations. 
Since these solutions describe the gravitational field of these BHs \cite{Kurmanov:2021uqv,Boshkayev:2021wns,Boshkayev:2020kle},  and/or the field outside massive compact objects, such as white dwarfs, neutron stars \cite{Shapiro:1983du}, and more exotic ones\footnote{The possible existence of exotic compact objects is not excluded, as BH candidates alone cannot explain observations, i.e., they are still not able to probe the geometry in proximity  to astrophysical sources. Remarkable examples are gravitational waves emitted from the inspiral of binary BHs \cite{Nakama:2020vtw}, star motion  near the galactic center \cite{Tito:2018wiu},  supermassive BH shadows  \cite{EventHorizonTelescope:2019dse}, etc.}, one needs to fix the symmetry underlying the corresponding spacetime \cite{Boshkayev:2021chc}.

Modelling the compact object at the center of galaxies surrounded by DM is not a simple task \cite{Jusufi:2019nrn}. The DM halo is quite completely unknown and its functional behaviour with respect to the radial coordinate $r$ can be determined only through simulations \cite{Lapi:2018nuq}, indirect strategies of modelling \cite{Hurst:2014uda}, or Monte Carlo analyses \cite{Hague:2013aqa}, giving rise to a wide number of compelling DM models \cite{DelPopolo:2009xj}.

We can therefore start from the simplest basic demands quite accepted in the literature, among which the existence of BHs 
at the center of galaxies, as suggested by emission lines of quasars \cite{Davies:2011pd},
with a spherically-distributed DM, modelled through suitable versions of the profile density. 

In view of the above considerations, it would be possible to estimate the radiative flux emitted by the accretion 
disk, along with its specific spectral luminosity distribution, commonly observed at infinity. The corresponding 
information 
one obtains would be enough to characterize both the BH central objects in terms of mass and fundamental properties and the 
nature of DM itself \cite{Bertone:2018krk}. 

As stated above, choosing the density profile of the DM envelope that will suffice to approximate a reasonable matter distribution turns out to be essential in order to compare the expectations from the presence of PBHs with respect to a 
smooth DM contribution.

\subsection{The sphere envelope distribution and system configuration}

Recent computations of the accretion disk luminosity 
\cite{Boshkayev:2021wns,Boshkayev:2020kle,Boshkayev:2021chc} have shown that a suitable choice for the DM 
distribution is offered by the exponential sphere density (ESD) \cite{Sofue:2008wt,Sofue:2008wu}:
\begin{equation}
\rho_\text{ESD}(r) = \rho_0\,  e^{-\frac{r}{r_0}}\,, \quad r\geq r_b\,.
\label{eq:Sofue_density}
\end{equation}
where $\rho_0$ is the central density value at $r=0$, while  $r_0$ is the scale radius after which the DM effects become 
negligible. 
In such a scheme, $r_b$ represents the inner edge of the DM envelope, i.e., the boundary separating the interior vacuum from the exterior region. The interior region is fully described by the BH gravitational field that dominates over the other matter contributions.
Although there is no practical indication about the size at which the inner edge of the DM envelope should be placed, it is reasonable to consider $r_b$ to be greater than the BH event horizon $2 M_{\text{BH}}$. Moreover, it appears as well clear that the DM envelope cannot reach $r=0$, so we can assume a constant $\rho_0$ not  interfering with the BH configuration field.

In this study, we shall take the standard approach described by Eq. \eqref{eq:Sofue_density} as a reference indicator to 
investigate physical observables derived from assuming the DM envelope made of PBHs distributed according to Eq. 
\eqref{eq:PBH_density}.
Taking into account the simplest spherical symmetry on the DM envelope, one immediately finds the mass of the DM 
constituent as
\begin{equation}
M_\text{DM}(r)=\int_{r_b}^r 4 \pi  r'^2 \rho (r') \, dr'\,.
\label{eq:integrale massa}
\end{equation}

Neglecting any other contribution, the total mass profile of the system composed by the presence of internal BH in addition 
to the DM envelope can be thus written as follows:
\begin{equation}\label{eq:massprof}
 M(r)=M_\text{BH}+M_\text{DM}(r)\,,\hspace{0.52cm} r_b \leq r \leq r_s,
\end{equation}
where $r_s$ can be identified with the surface radius of the DM envelope. The latter quantity will enter as a free 
parameter in the TOV equations once we compute them to obtain the accretion disk luminosity. Throughout the paper, we use 
geometrized units such that $G=c=1$.

Therefore, applying Eq. \eqref{eq:integrale massa} to the PBH density profile \eqref{eq:PBH_density} yields
\begin{equation}
M_\text{DM}^\text{(PBH)}(r)=\dfrac{M_\ast}{R_\ast^{3/5}}\left(r^{3/5}-r_b^{3/5}\right), \quad r>r_b\,.
\end{equation}

On the other hand, for the reference ESD model \eqref{eq:Sofue_density} we obtain 
\begin{equation}
M_\text{DM}^\text{(ESD)}(r)=6M_0e^{-\frac{r_b}{r_0}}\Big[g(r_b)-e^{\frac{r_b-r}{r_0}}g(r)\Big],\quad r>r_b\,,
\end{equation}
where $M_0= \frac{4}{3}\pi r_0^3\rho_0$ and $g(r)\equiv 1+\frac{r}{r_0}+\frac{r^2}{2r_0^2}$.

\subsection{Relativistic effects and symmetry of the system}

Adopting general relativity and spherical symmetry, in \cite{Novikov-Thorne,Page:1974he} astrophysical BHs were used to explain 
the features of their observed spectrum\footnote{Alternative approaches have been carried out with BHs in vacuum, for 
instance adopting the Kerr metric \cite{Harko:2009gc,Harko:2009xf}. Recently, the interest is shifted to 
accretion disks in a geometry that departs from the Kerr line element also \cite{ 
Bambi:2011jq,Bambi:2013hza}.}.
Relativistic compact objects are naively portrayed as fulfilling  symmetry conditions and numerical matching between the 
exterior and interior solutions, by  solving hydrodynamic equilibrium equations for the matter contained in the interior 
\cite{Joshi:2013dva}. 
Without limiting to BHs, also compact and/or exotic stable  compact objects can be featured by virtue of the above picture, 
leading to the still open issue that such massive objects may be DM condensates \cite{Levkov:2018kau}. Intriguingly, 
depending on the DM nature, the above symmetry provides a non-vanishing pressure term, whose consequences can be 
investigated either in cosmological contexts (e.g.~\cite{Luongo:2018lgy}), or in relativistic astrophysics 
(e.g.~\cite{Arbey:2021gdg}).
This standard approach, however, is limited due to the appearance of singularities where general relativity fails to be 
predictive \cite{Malafarina:2017csn},
and quantum gravity may be adopted as large fields are involved in the scheme\footnote{Remarkably,  vacuum, static, and 
axially symmetric solutions, known as Weyl's class \cite{Weyl:1917gp,Weyl:1919fi} show  curvature singularities at the infinitely 
redshifted surface, i.e. at the BH horizons  \cite{Hernandez-Pastora:2011rbo}. Even though singularities occur, such a class of solutions describes the exterior field of static and axially symmetric exotic compact objects, if one assumes a 
corresponding boundary placed  at a distance outside the singularity.} \cite{Harko:2011kw}.

Therefore, to study the properties of the system modelled as we depicted above, we consider the metric
\begin{equation}
d s^2=e^{\eta(r)} d t^2 - e^{\lambda(r)} d r^2 - r^2 \left(d \theta^2 + \sin^2 \theta\, d\varphi^2\right),
\label{metric}
\end{equation}
where $(t,r,\theta,\varphi)$ are the time and spherical coordinates, $\eta(r) $ and $\lambda(r)$ are the sought metric 
functions. 

From \eqref{metric} we thus find the TOV equations for the DM envelope:
\begin{align}
\frac{dP(r)}{d r}&=-\left(\rho(r) +P(r)\right) \frac{M(r)+4 \pi r^3 P(r)}{r(r-2 M(r))}\,,\label{eq:TOV1}\\
\frac{d \eta(r)}{dr}&=-\frac{2}{\rho(r)+P(r)}\frac{d P(r)}{d r}\, ,\label{eq:TOV2}
\end{align}
where $P(r)$ and  $\rho(r)$ are the DM pressure and density, respectively, and the mass $M(r)$ is given by 
Eq.~\eqref{eq:massprof}. The other unknown coefficient of the metric \eqref{metric}, i.e. $g_{rr}$, is well known to satisfy
\begin{equation}
   e^{\lambda (r)}=\left(1-\frac{2M(r)}{r}\right)^{-1}. 
\end{equation}

To summarize, in our approach we consider the central object, surrounded by an external cloud, and the accretion disk.
The latter lies on the equatorial plane and, consequently, its symmetry breaks the one provided by  the spherical cloud. In this respect, the accretion disk appears fully non-spherical, prompting instead an axisymmetric configuration, made up of the matter particles that fuel the accretor itself. 
Indeed, the accretion disk emits radiation, whereas the DM cloud influences the geodesics of the particles moving on the accretion disk. 
Therefore, our analysis is based on two main assumptions: 
particles approximately move along geodesics since the DM influence is negligible;
the accretion disk is smaller than the $10\%$ of the object that generates it, as argued in \cite{Bambi:2014koa}.

\subsection{Boundary conditions}

Our system is described by a static metric under the  spherical layer  of DM that represents the envelope around the BH. 
Naturally, under the above hypothesis, the matching between the inner part with the outer one needs peculiar requirements. 
Matter should smoothly fill from  the boundary $r_b$ up to an external Schwarzschild solution  \cite{Fayos:1996gw}.
This fact can be fulfilled by simply requiring the pressure to vanish at $r=r_s$ and continuity on the mass density. Some 
more details on this point are raised in Appendix \ref{sec:matching}.

In view of the aforesaid, at the outer boundary of the DM envelope, one has to impose conditions over the pressure
\begin{equation}
P(r_s)=0\,, 
\label{prs}
\end{equation}
stating that the DM pressure vanishes at the most external surface of the galaxy. Moreover, the mass of the vacuum 
exterior will be $M(r_s)$.   

In addition, as boundary for $\eta(r)$, we will consider
\begin{equation}
\eta(r_b)=\ln{\left(1-\frac{r_g}{r_b}\right)}\,.
\label{rhoetarb}
\end{equation}

It is worth noticing that, if one asks for a smooth matching between \eqref{metric} and a Schwarzschild interior BH, then 
$P(r)$ should vanish at $r=r_b$ too. 

However, the DM density is not zero at $r=r_b$, suggesting the matching condition over the pressure cannot give $P(r_b)=0$. 
To clarify this point, we can invoke the Newtonian limit. There,  non-continuous matching between interior and exterior is 
not possible and, consequently,  fulfilling the Newtonian suggestion, $P(r_b)\neq0$ even in our relativistic case. This 
prevents an un-physical jump at the boundary, i.e., the eventual pressure discontinuity from zero to a non-zero value of 
$P(r)$, and avoids \emph{de facto} disagreement with current observations that seem not to indicate thermodynamic 
discontinuities in spiral galaxies. 

This problem can be healed by assuming the presence of a massive surface layer at $r_b$ 
\cite{Israel:1966rt}. Thus, as the corresponding pressure cannot be zero, we assume $P(r_b)\neq0$.

The TOV equations do not possess analytical solutions, so we need to proceed numerically. 
This task in principle is not straightforward to implement, as the numerical solutions fail to converge for particular sets 
of boundaries, due to the superlinear behavior with respect to $P(r)$ in \eqref{eq:TOV1}. Motivated by the study of 
\cite{Boshkayev:2020kle}, we set the following indicative numerical values: $M_\text{BH}=5\times 10^8 M_\odot\approx$ 4.933 
AU\footnote{1 astronomical unit (AU) $\approx 8.9 \times 10^{23}M_\odot$ pc$^{-3}$.}, $r_b=5.5\, M_\odot\approx 27.133$ AU, 
$r_0=10$ AU, $r_s=220$ AU and $\rho_0=10^{-5}$ AU$^{-2}$. Moreover, as for the PBH distribution, we assume $h_0=0.7$ and 
$z\simeq 0$ in our computations.

\section{Accretion disk luminosity}\label{sezione4}

In our treatment, particles follow circular geodesics in the equatorial plane $\theta=\frac{\pi}{2}$. We also assume that 
PBHs behave as identical particles with the same properties. As the metric is static, the energy is conserved, and the disk 
will be characterized by particles with specific angular momentum and angular velocity, which depend on the distance from 
the central BH.

To determine the accretion disk luminosity, we need to define the radiative flux, i.e. the energy radiated per unit area 
per unit time, which is emitted by the accretion disk \cite{Joshi:2011zm}: 
\begin{equation}\label{eq:flux}
    {\cal F} (r)=- \frac{\dot{m}}{4\pi \sqrt{g}} \frac{\omega_{,r}}{(\mathcal{E}-\omega \ell)^2} \int_{r_i}^{r}  dr'\, 
(\mathcal{E}-\omega \ell) \ell_{,r'}\,,
\end{equation}
where the dot indicates the time derivative, and the comma denotes the derivative with respect to the subsequent specified 
variable. Here, $\dot{m}$ is the mass accretion rate, $g$ is the determinant of the metric tensor, $\mathcal{E}$ is the 
energy per unit mass, while $\omega$ and $\ell$ are the orbital angular velocity and momentum per unit mass, respectively. 
Furthermore, $r_i$ is the radius of the inner edge of the disk, obtained from the condition $\ell_{,r}=0$. 

Our investigation does not involve observable data, but rather we shall simulate the shapes of luminosity related to how DM influences the accretion disk properties. The corresponding model is therefore chosen to be the simplest one, i.e. Novikov-Thorne, that appears physically well-motivated, as previously stressed.

Furthermore, we assume that DM is approximately static around BHs. 
Clearly, this happens since the accretion disk rate is shorter than DM falling into the BH itself. The approximation appears well motivated since we are claiming that DM does not contribute to BH masses, which is reasonable if we focus on short time intervals. Indeed, after long periods, this approximation breaks down and so to avoid possible misleading considerations, we take the mass rate, namely $\dot m$, roughly constant, while the accretion disk produces luminosity and flux.

In the equatorial plane, we have
\begin{align}
\omega(r)&= \sqrt{- \frac{g_{tt,r}}{g_{\varphi\varphi,r}}}\,, \\
\ell(r)&=-u_{\varphi}=-u^{\varphi}g_{\varphi\varphi} =-\omega\, u^{t}g_{\varphi\varphi}\,, \\
\mathcal{E}(r)&=u_{t}=u^{t}g_{tt}, \\
u^{t}(r)&=\frac{1}{\sqrt{g_{tt} + \omega^{2} g_{\varphi\varphi}}}\,.
\end{align}
where $u_{t}$ and $u_{\varphi}$ are the covariant time and angular components of the four-velocity, respectively.
In order to better display our results, it turns convenient to work with the following dimensionless functions: 
$\tilde{\omega}(r)\equiv M_s\,\omega(r)$ and $\tilde{\ell}(r)\equiv \ell(r)/M_s$, where $M_s=M(r_s)$.

Another useful quantity to consider, rather than the indirectly observable flux, is the luminosity reaching the observer at infinity, ${\cal L}_\infty$, defined through \cite{Novikov-Thorne,Page:1974he}
\begin{equation}\label{eq:difflum}
 \frac{d{\cal L}_{\infty}}{d\ln{r}}=4\pi r \sqrt{g}\,\mathcal{E}\,{\cal F}(r)\,.
\end{equation}

Assuming that the accretion disk gas behaves as a perfect blackbody, one then finds the spectral luminosity observed at 
infinity \cite{Joshi:2013dva}:
\begin{equation}\label{eq:speclum}
    \nu{\cal L_{\nu,\infty}}=\frac{15}{\pi^4}\int_{r_i}^{\infty}d\ln {r}\, \left(\frac{d{\cal L_{\infty}}}{d 
\ln{r}}\right)\frac{(u^t y)^4/\tilde{\mathcal{F}}}{e^{u^t y/\tilde{\mathcal{F}}^{1/4}}-1}\,,
\end{equation}
where $y\equiv\frac{h\nu}{k_BT_{c}}$, with $h$ being the Planck constant, $T_c$ the characteristic temperature, $\nu$ the 
emitted radiation frequency and $k_B$ the Boltzmann constant. We have also defined  $\tilde{\mathcal{F}}(r)=M(r_s)^2{\cal 
F}(r)$. 

Hence, using Eq.~\eqref{eq:difflum}, we can write Eq.~\eqref{eq:speclum} as
\begin{equation}
\nu{\cal L_{\nu,\infty}}=\frac{60}{\pi^3}\int_{r_i}^{\infty} dr\, \frac{\sqrt{g}\, \mathcal{E} }{M_s^2}\frac{(u^t 
y)^4}{e^{u^t y/\tilde{\mathcal{F}}^{1/4}}-1}\,.
\end{equation}

Concerning the use of accretion disk luminosity, we want to stress that emission lines from the central regions give hints about the strong gravity regime and, consequently, provide information on the accretor itself. Among the ones in the 
X-ray spectrum, the $K\alpha$ line at $6.4$ keV is the most relevant \cite{Reynolds:1998ie}. Due to its broadening features, such 
a line represents an outstanding probe for the geometry around central BHs and/or in general around accretors 
\cite{Bambi:2012at}.
Clearly, the presence of a PBH envelope will affect also the $K\alpha$ line, since a matter distribution close to BHs 
alters the broadening of the line. This will be thus a source of few uncertainties in measuring the BH mass.

\section{Theoretical results}\label{sezione5}

\begin{figure}
\includegraphics[width=3.3 in]{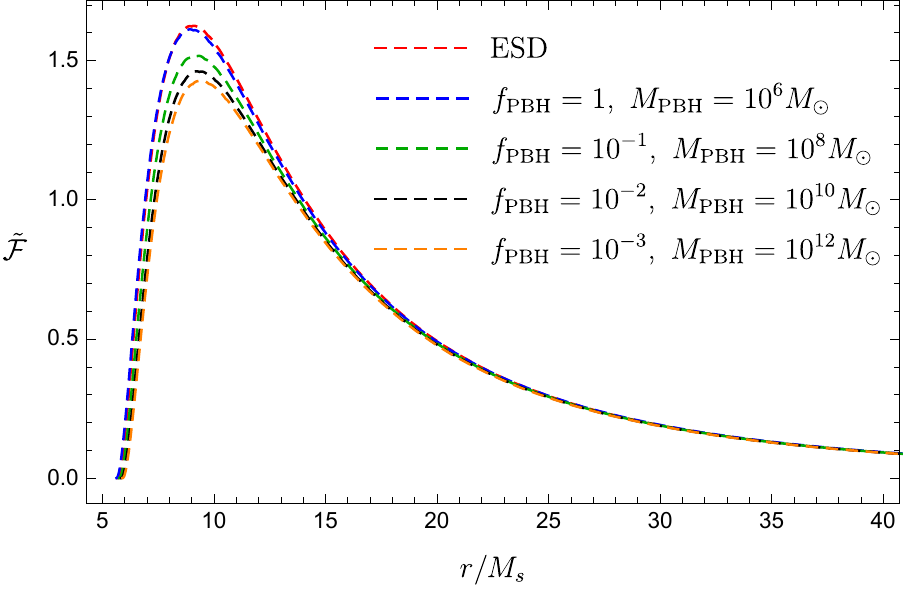}
\caption{Radiative flux of the accretion disk normalized to $10^{-5}$ as a function of the radial distance for different 
PBH fractions and masses, compared to the prediction of the ESD model.}
\label{fig:F}
\end{figure}

\begin{figure}
\includegraphics[width=3.3 in]{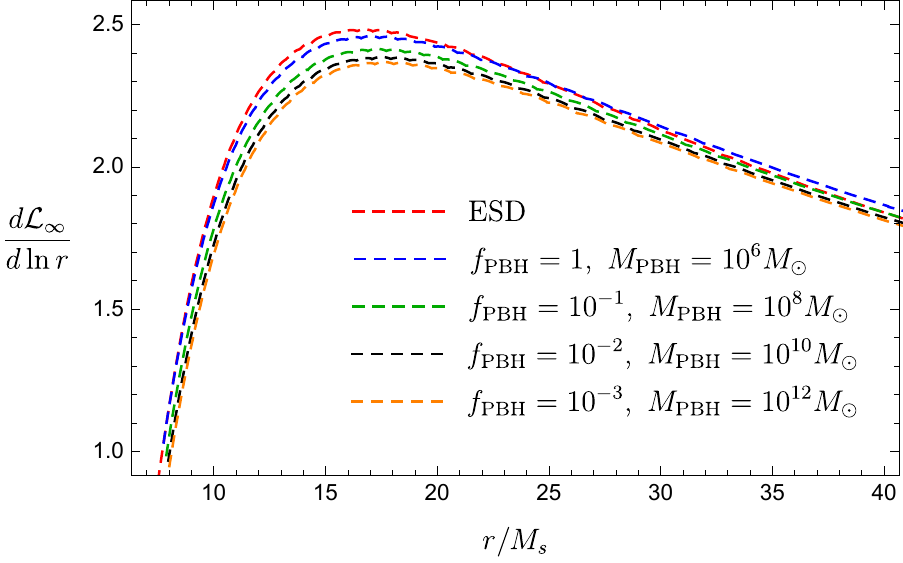}
\caption{Differential luminosity of the accretion disk normalized to $10^{-2}$ as a function of the radial distance for 
different PBH fractions and masses, compared to the prediction of the ESD model.}
\label{fig:dL}
\end{figure}

\begin{figure*}
\centering
\subfigure
{\includegraphics[width=3.3 in]{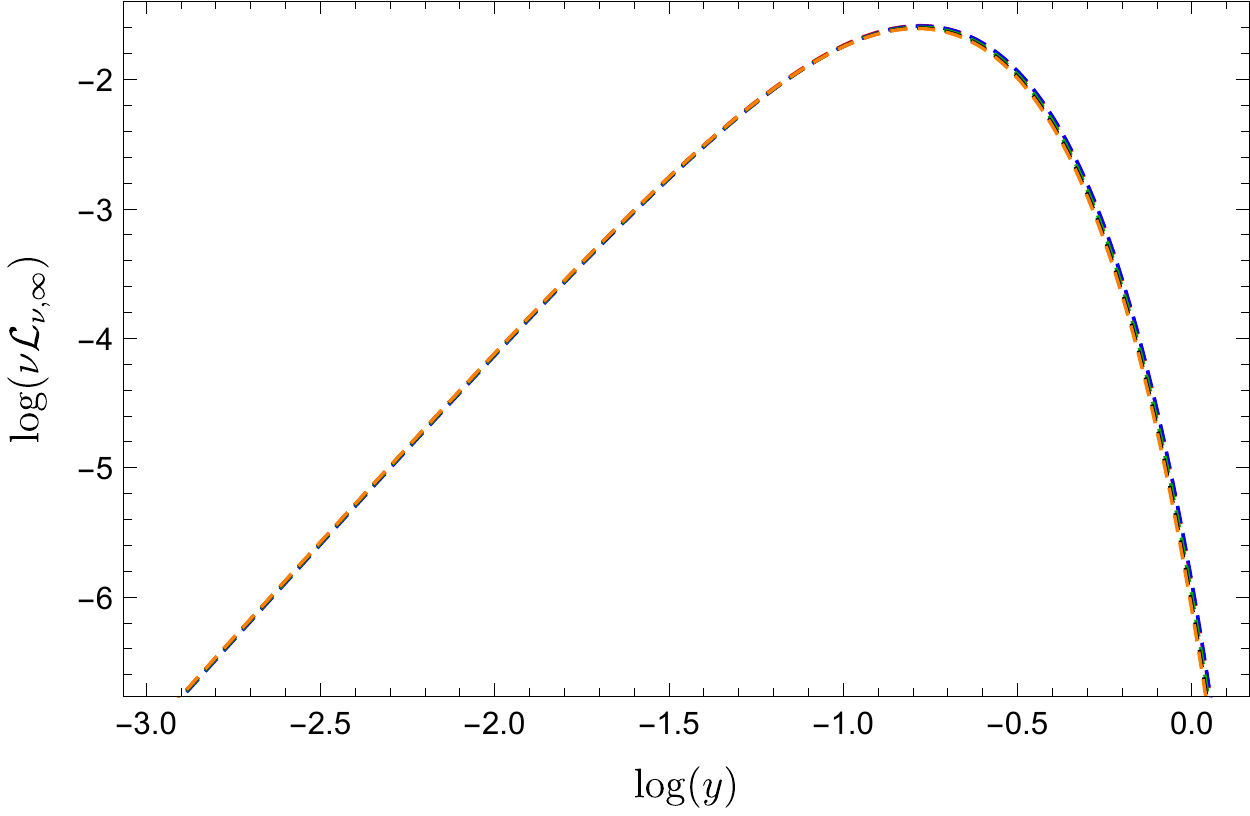}}
\hfill
\subfigure
{\includegraphics[width=3.4in]{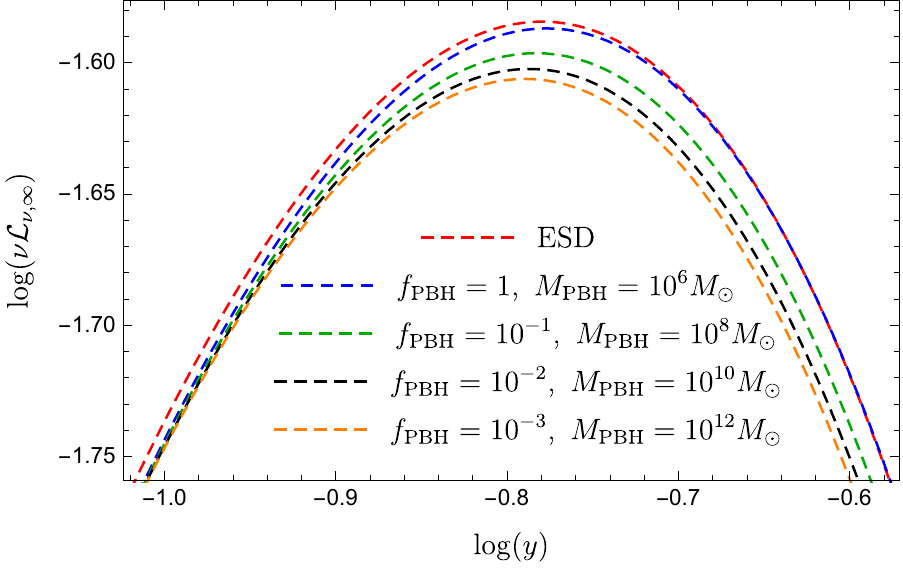}}
\caption{Spectral luminosity as a function of the radiation frequency for different PBH fractions and masses, compared to 
the prediction of the ESD model.  The right panel shows a zoom on the corresponding peak.}
\label{fig:spectral}
\end{figure*}

We performed our numerical computation by focussing on a set of four values of PBH fractions, 
$f_\text{PBH}=(10^{-3},10^{-2},10^{-1},1)$, corresponding to which a broad range of PBH masses has been considered. From 
the 
solutions of the TOV equations, we found that the combinations of PBH fractions and masses that are consistent with 
theoretical expectations are those shown in Figs. \eqref{fig:F}-\eqref{fig:spectral}. In fact, all other choices of 
$M_\text{PBH}$ for each $f_\text{PBH}$ value are characterized by either integration issues or observational curves that 
are 
severely discrepant with respect to the reference ESD model. 

From Figs.~\eqref{fig:F} and \eqref{fig:dL} we note that the flux and luminosity peaks tend to increase as the more PBH 
fraction is considered. 
Interestingly, our results suggest that lower PBH fractions prefer higher PBH masses. Also, it is worth noting the 
remarkable case offered by the combination $f_\text{PBH}=1$ and $M_\text{PBH}=10^6 M_\odot$, which is capable of 
reproducing 
very closely the features of the ESD model. 

The spectral luminosity shown in the left panel of Fig. \eqref{fig:spectral} apparently does not provide remarkable 
indications. However, the closer focus around the peak region in the right panel of Fig. \eqref{fig:spectral} indicates 
slight departures among the curves mostly evident for $\log(y) \gtrsim -1$, with a relative change in the spectral 
luminosity $\lesssim$ 5\%. Again, the result that better matches the ESD expectation is achieved for $f_\text{PBH}=1$ and 
$M_\text{PBH}=10^6 M_\odot$.

The kinematic quantities involved in our picture (see Appendix \ref{sec:kin}) confirm what we stated above. In particular, 
Fig. \eqref{fig:E} shows that the energy tends to decrease for lower PBH densities as the source to fuel DM. Also, the matching 
between $f_\text{PBH}=1$ and $M_\text{PBH}=10^6 M_\odot$ curve with the ESD prediction is clearly the best one that we 
obtain at all radii. 

On the other hand, the orbital angular momentum (see Fig. \eqref{fig:L}) curves are hardly appreciable 
at large distances, while the angular velocity curves are practically indistinguishable among them (see Fig.~\eqref{fig:omega}). This fact represents a limitation of the theoretical setup that one should overcome in order to detect 
any significant modification induced by the presence of PBHs.

Finally, we can check the consistency of our results in light of current observations. Specifically, the most recent constraints on the DM abundance by the Planck Collaboration \cite{Planck:2018vyg} suggest 
\begin{equation}\label{ilnu}
    \overline \rho_\text{DM}\simeq 3.2\cdot 10^{-8}M_\odot/\text{pc}^3\,.
\end{equation}
Comparing the latter to $\overline \rho_\text{PBH}=\overline n_\text{PBH}M_\text{PBH}$ by means of Eq.~\eqref{eq:n_medio}, within our assumptions, we obtain $f_\text{PBH}\simeq 1$. This remarkably matches our theoretical prediction, implying that the hypothesis of PBHs as DM candidates is compatible with the current understanding of DM properties.
In particular, Eq.~\eqref{ilnu} shows that there are large enough overdensities capable of clustering and forming the expected halos, in analogy with the hypothesis of assuming DM to be made of particles that are not predicted by the standard model of particle physics \cite{ParticleDataGroup:2022pth}.

\section{Outlook and perspectives}\label{sezione6}

In this study, we considered the hypothesis that PBHs constitute a fraction of the DM envelope surrounding a supermassive BH placed at the center of a generic spiral galaxy. 

To describe the whole system, we fixed the geometry of the galaxy according to spherical symmetry, modelling the bulge with 
a Schwarzschild solution and the rest by means of a TOV spacetime. Boundary conditions on the galaxy configuration have 
been 
discussed, together with the strategy to solve the TOV equations.

We thus analyzed the physical properties of the accretion disk, investigating how it gets modified due to the presence of 
PBHs. To this purpose, we modelled a suitable PBH density profile taking into account the two-point correlation function of 
PBH halos. 
Therefore, through the Novikov-Thorne luminosity, we evaluated the flux, energy, angular velocity and momentum of the 
accretion disk under the assumption of negligible mass loss.

Bounds on the PBH abundance and mass were inferred by comparing our theoretical predictions with the outcomes provided by 
the ESD profile. We found that the most suitable PBH mass range is  $(10^6-10^{12})M_\odot$ for 
$f_\text{PBH}\in[10^{-3},1]$, with the best-matching configuration given by $f_\text{PBH}=1$ and $M_\text{PBH}=10^6 
M_\odot$.

From our results, we can conclude that including PBHs would clearly modify the spectral properties of a galaxy. The 
corresponding accretion disk luminosity seems to favour large fractions of PBHs with corresponding masses that are the 
smallest ones within the range of all suitable possibilities. Clearly, our theoretical findings rely on the initial 
settings 
and depend on the galaxy structure model, and the numerical solutions of the TOV equations. Likely, more refined and 
realistic analyses, based on specific galaxies, may be able to provide us with more accurate bounds. Nevertheless, the 
behaviours of all curves show how we expect the PBHs would modify observations related to the accretion disk luminosity. 

In this regard, our approach may represent a tool to restrict viable windows of abundances and masses of PBHs, even if based on a toy-model scenario relying on the use of the Novikov-Thorne luminosity. Alternative scenarios requiring additional efforts will be investigated in future works.

\begin{acknowledgments}
R.D. acknowledges the support of INFN (\emph{iniziativa specifica} QGSKY). O.L. acknowledges the Ministry of Education and Science of the Republic of Kazakhstan, Grant: IRN AP08052311. The authors are grateful to K. Boshkayev, T. Konysbayev and E. Kurmanov for useful discussions.
The authors would like to express their gratitude to the anonymous reviewer for her/his valuable comments and helpful suggestions.
\end{acknowledgments}

\appendix

\section{Kinematic quantities}\label{sec:kin}

In this appendix, we display the behaviour of the kinematic quantities related to the accretion disk as a function of the 
radial distance from the central BH.

\begin{figure}
\includegraphics[width=3.3 in]{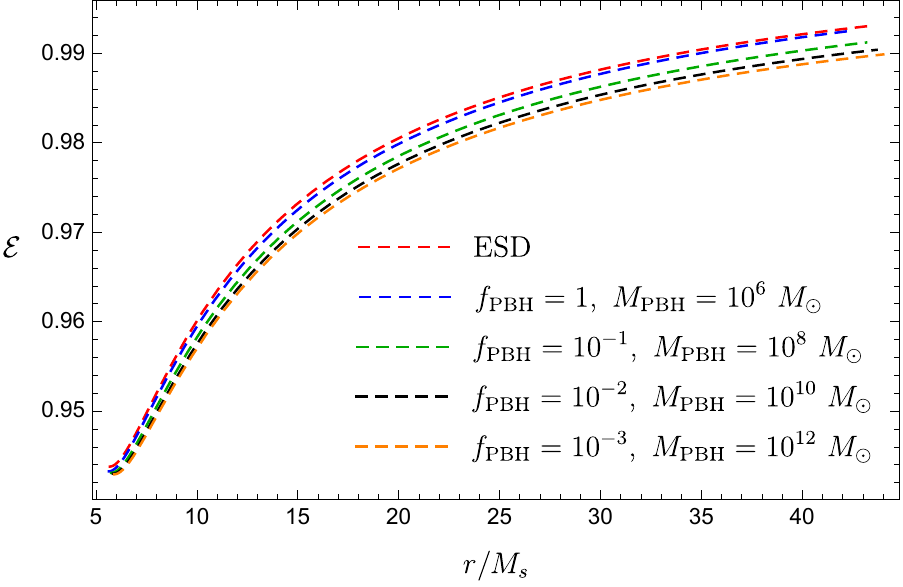}
\caption{Energy as a function of the radial distance for different PBH fractions and masses, compared to the prediction of 
the ESD model.}
\label{fig:E}
\end{figure}

\begin{figure}
\includegraphics[width=3.3 in]{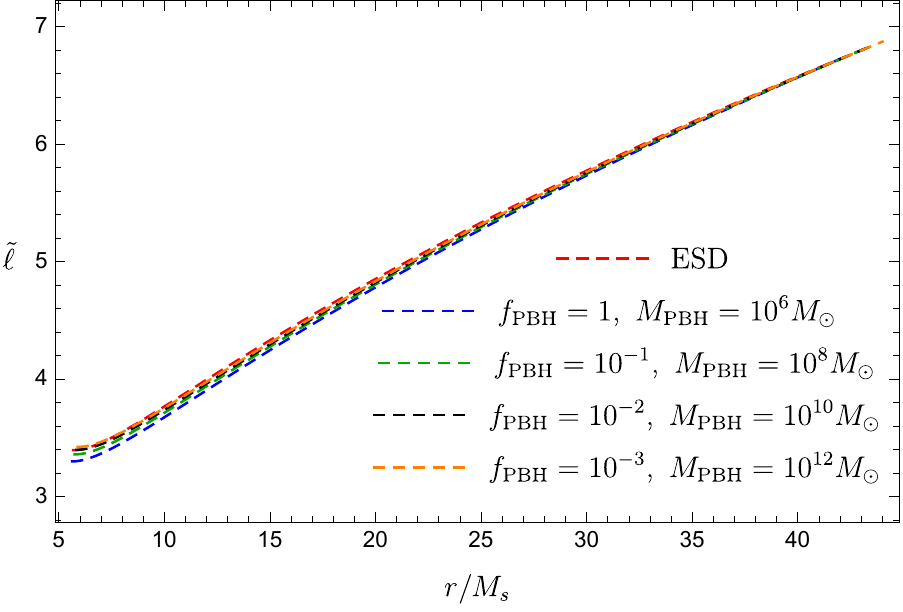}
\caption{Angular momentum as a function of the radial distance for different PBH fractions and masses, compared to the 
prediction of the ESD model.}
\label{fig:L}
\end{figure}

\begin{figure}
\includegraphics[width=3.3 in]{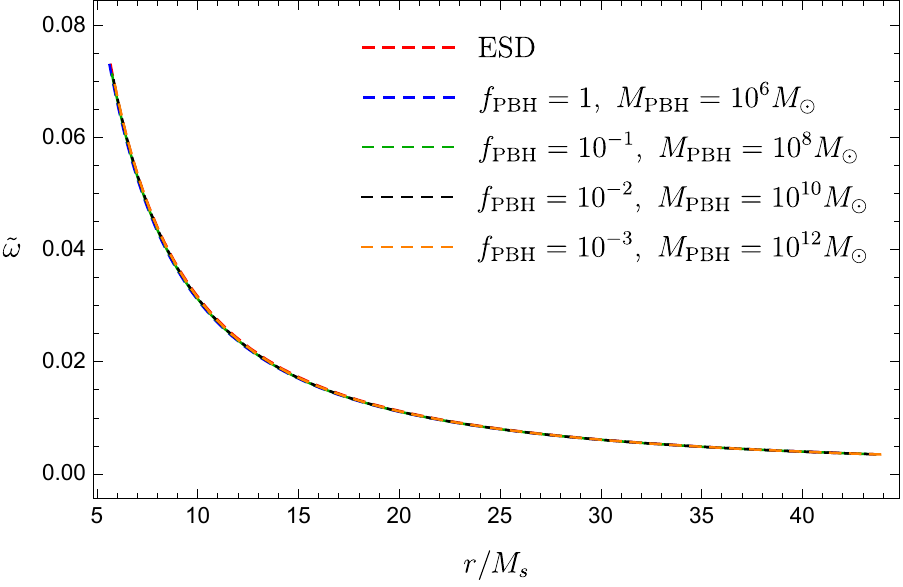}
\caption{Angular velocity as a function of the radial distance for different PBH fractions and masses, compared to the 
prediction of the ESD model. The right panel shows a zoom on large distances.}
\label{fig:omega}
\end{figure}

\section{Matching with the external solution}\label{sec:matching}

In some recent studies \cite{Boshkayev:2020kle,Boshkayev:2021chc,Boshkayev:2021wns}  it was pointed out that the vanishing 
pressure at the surface radius, $r_s$, requires the condition $\eta(r_s)+\lambda(r_s)=0$ to hold. 
However, the numerical solution $\eta_n(r_s)$ emerging from the TOV equations does not satisfy such constraint and, thus, 
a possible choice is to redefine the function $\eta$ as 
\begin{equation}
\eta_{r}(r)=\eta_{n}(r)-\left[\eta_{n}(r_s)-\ln{\left(1-\frac{2M(r_s)}{r_s}\right)}\right]\frac{r-r_b}{r_s-r_b}\,.
\end{equation}

The above function, in fact, fulfills the boundary conditions $
    e^{\eta (r)}=                  1-\frac{r_g}{r}$ at $r=r_b$ and $e^{\eta (r)}=1-\frac{2 M(r_s)}{r}$ at $r= r_s$.
However, this strategy suffers from two main shortcomings. 
Indeed, redefining $\eta(r)$ implies modifications of the pressure and density, which would not be anymore solutions of the 
original TOV equations.
Actually, that choice for $\eta$  is not strictly required to satisfy Israel's conditions for a smooth matching with a 
spherical vacuum solution \cite{Israel:1966rt}. To guarantee, in fact, the matching between the 
interior and exterior galaxy regions, one needs only to assume $P(r_s)=0$ and the continuity of mass (see e.g. 
\cite{Giambo:2005se}). For these reasons, in the present paper, we consider $\eta$ as obtained from the TOV solutions, without any additional reparametrization.

\end{document}